\documentclass[onecolumn,notitlepage,superscriptaddress,longbibliography,showpacs,prd,aps]{revtex4-1}

\usepackage{amsmath}
\usepackage{amsfonts}
\usepackage{amssymb}
\usepackage{color}
\usepackage{graphicx}
\usepackage{cancel}
\usepackage{dsfont}
\usepackage{bm}
\usepackage{maybemath}

\newcommand{\vecrhoprime}{\vec{\rho}{\,'}}

\newcommand{\dd}{\mathrm{d}}
\newcommand{\ee}{\mathrm{e}}
\newcommand{\ii}{\mathrm{i}}

\begin{document}

\title{Green Function of the Poisson Equation: $\bm{D=2,3,4}$}

\author{U.~D.~Jentschura}

\affiliation{Department of Physics,
Missouri University of Science and Technology,
Rolla, Missouri 65409, USA}

\author{J. Sapirstein}

\affiliation{Department of Physics, University of Notre Dame, 
Notre Dame, Indiana 46556, USA}

\begin{abstract}
We study the Green function of the Poisson equation in 
two, three and four dimensions.
The solution $g$ of the equation 
$\vec\nabla^2 g(\vec x - \vec x') = \delta^{(D)}(\vec x - \vec x')$,
where $\vec x$ and $\vec x'$ are $D$-dimensional
position vectors, is customarily expanded into 
radial and angular coordinates.
For the two-dimensional case ($D=2$), we
find a subtle interplay of the 
necessarily introduced scale $L$ 
with the radial component of zero magnetic quantum number.
For $D=3$, the well-known expressions are briefly recalled;
this is done in order to highlight the analogy 
with the four-dimensional case, where
we uncover analogies of the four-dimensional spherical harmonics 
with the familiar three-dimensional case.
Remarks on the $SO(4)$ symmetry of the hydrogen atom complete
the investigations.
\end{abstract}


\maketitle

%
%
\section{Introduction}

Solutions of the equation
\begin{equation}
\label{master}
\vec\nabla^2 g(\vec x - \vec x') = \delta^{(D)}(\vec x - \vec x')
\end{equation}
enter a myriad of physical problems, from the elementary Coulomb problem 
in electrostatics ($D=3$),
to the attraction among vortices in two-dimensional systems ($D=2$),
and on to the four-dimensional formulation of the 
hydrogen Green function ($D=4$, see Ref.~\cite{Sc1964}).
Here, we shall attempt to provide a unified treatment 
of the radial and angular decompositions of the 
two-, three- and four-dimensional Green functions,
which are solutions to Eq.~\eqref{master}.

In $D=2$, a scale has to be introduced, which 
corresponds to a physically irrelevant 
overall constant term, while 
in $D=3$, the formulas are very familiar
(see Refs.~\cite{Ja1998,Je2017book}).
In $D = 4$, we attempt to reveal a 
structure of (re-)defined associated ultraspherical 
polynomials (Gegenbauer polynomials), 
which highlights analogies to the associated Legendre functions 
that enter the case $D=3$.

%
%
\section{Two--Dimensional Case}

Spherical coordinates in two dimensions have a 
cylindrical symmetry; hence, for definiteness, we 
denote the two-dimensional position vectors as
$\vec\rho$ and $\vecrhoprime$, and their 
moduli as $\rho = | \vec \rho |$ and $\rho' = | \vecrhoprime |$.
The Green function solution of the Poisson equation,
\begin{equation}
\label{gf}
\vec\nabla^2 g(\vec \rho - \vecrhoprime) =
\delta^{(2)}(\vec \rho - \vecrhoprime) \,,
\quad
g(\vec \rho - \vecrhoprime) =
\frac{1}{2 \pi} \, \ln\left( \frac{| \vec\rho - \vecrhoprime |}{L} \right),
\end{equation}
introduces a scale $L$, which ensures that the 
argument of the natural logarithm is dimensionless.
In terms of the Green function, the scale $L$ 
adds nothing but an overall constant term,
\begin{equation}
g \to g - \frac{1}{2\pi} \, \ln \, L. 
\end{equation}
In order to show that 
$g$ fulfills the Poisson equation, one specializes 
the divergence theorem to an infinitesimal area $A$.
For example, $A$ might be chosen as the 
inner area of a circle of infinitesimal 
radius $\epsilon$, about the center $\vecrhoprime$.
With $\partial A$ denoting the boundary of $A$, 
i.e, the circle of radius $\epsilon$ about $\vecrhoprime$, one 
must have
\begin{equation}
\int_{\partial A} \vec\nabla g(\vec\rho - \vecrhoprime) \cdot \dd \vec S 
= \int_A \vec\nabla^2 g(\vec\rho - \vecrhoprime) \, \dd^2 \rho
= \int_A \delta^{(2)}(\vec\rho - \vecrhoprime) \, \dd^2 \rho  = 1 \,.
\end{equation}
Using the formula~\eqref{gf} and the radial component of the 
gradient operator in two-dimensional coordinates, one verifies that 
indeed,
\begin{equation}
\int_{\partial A} \vec\nabla g(\vec\rho - \vecrhoprime) \cdot \dd \vec S 
= \int_0^{2 \pi} 
\left. \frac{\partial}{\partial \rho} g(\vec\rho - \vecrhoprime) 
\right|_{ | \vec\rho - \vecrhoprime | = \epsilon} \, 
\epsilon  \, \dd \varphi 
\nonumber\\[0.1133ex]
= \int \frac{1}{2 \pi \epsilon} \, \epsilon \, \dd \varphi = 1 \,,
\end{equation}
independent of $\epsilon$.

Let us now turn to the angular-momentum decomposition of 
Eq.~\eqref{gf}.
The spherical representation of the two-dimensional 
Dirac-$\delta$ is 
\begin{equation}
\delta^{(2)}(\vec \rho - \vecrhoprime) = 
\frac{1}{\rho} \, \delta(\rho - \rho') \, 
\delta(\varphi - \varphi') \,.
\end{equation}
%
An appropriate {\em ansatz} for the Green function is
\begin{equation}
g(\vec \rho - \vecrhoprime) =
\sum_{m = -\infty}^{\infty} f_m(\rho, \rho', \varphi') \; \ee^{\ii m \varphi} \,.
\end{equation}
The two-dimensional representation of the Laplacian is 
\begin{equation}
\vec\nabla^2 =
\frac{\partial^2}{\partial \rho^2} + 
\frac{1}{\rho} \, \frac{\partial}{\partial \rho} +
\frac{1}{\rho^2}
\frac{\partial^2}{\partial \varphi^2} \,.
\end{equation}
It acts on the Green function as follows,
\begin{equation}
\vec\nabla^2 g(\vec \rho - \vecrhoprime) 
= \sum_{m = -\infty}^\infty
\left( \frac{\partial^2}{\partial \rho^2} +
\frac{1}{\rho} \, \frac{\partial}{\partial \rho} -
\frac{m^2}{\rho^2} \right)
f_m(\rho, \rho', \varphi') \; \ee^{\ii m \varphi} 
= \frac{1}{\rho} \, \delta(\rho - \rho') \, 
\delta(\varphi - \varphi') \,.
\end{equation}
Now, one multiplies both sides with the factor
\begin{equation}
\frac{1}{2 \pi} \, \ee^{-\ii m' \varphi} \,,
\end{equation}
and integrates over $\dd \varphi$, resulting in 
the equation
\begin{equation}
\left( \frac{\partial^2}{\partial \rho^2} +
\frac{1}{\rho} \, \frac{\partial}{\partial \rho} -
\frac{m'^2}{\rho^2} \right) f_{m'}(\rho, \rho', \varphi')
= \frac{1}{2 \pi \rho} \, \delta(\rho - \rho') \, 
\ee^{-\ii m' \varphi'} \,.
\end{equation}
Setting
\begin{equation}
f_{m'}(\rho, \rho', \varphi') =
g_{m'}(\rho, \rho') \, \ee^{-\ii m' \varphi'} \,,
\end{equation}
and renaming $m' \to m$ after this operation,
one obtains the radial equation
%
%
\begin{equation}
\label{radeq}
\left( \frac{1}{\rho} \frac{\partial}{\partial \rho} 
\rho \, \frac{\partial}{\partial \rho} -
\frac{m^2}{\rho^2} \right) g_m(\rho, \rho')
= \frac{1}{2 \pi \rho} \, \delta(\rho - \rho') \,.
\end{equation}
%
%
%
Inspired by textbook treatments~\cite{Ja1998,Je2017book} of the 
three-dimensional Green function, 
one uses the following {\em ansatz} for nonzero $m$,
\begin{equation}
\label{derSTART}
g_m(\rho, \rho') = C \, \left( \frac{\rho_<}{\rho_>} \right)^{|m|} \,,
\qquad
m \neq 0 \,,
\end{equation}
where $\rho_< = \min(\rho, \rho')$ and $\rho_> = \max(\rho, \rho')$,
and integrates Eq.~\eqref{radeq} 
from $\rho = \rho' - \epsilon$ to $\rho = \rho' + \epsilon$,
\begin{equation}
\int\limits_{\rho = \rho' - \epsilon}^{\rho = \rho' + \epsilon}
\left( \frac{\partial}{\partial \rho}
\rho \, \frac{\partial}{\partial \rho} -
\frac{m^2}{\rho^2} \right) g_m(\rho, \rho') 
\, \dd \rho
= \int\limits_{\rho = \rho' - \epsilon}^{\rho = \rho' + \epsilon}
\frac{1}{2 \pi} \, \delta(\rho - \rho') \, \dd \rho \,.
\end{equation}
This results in the relation
\begin{equation}
\label{radcond}
\left. \left[ \rho \, \frac{\partial}{\partial \rho} 
g_m(\rho, \rho') \right]
\right|^{\rho = \rho' + \epsilon}_{\rho = \rho' - \epsilon}
= \frac{1}{2 \pi} \,,
\end{equation}
and amounts to the condition
\begin{equation}
C \, \left( 
\left. \left[ \rho \, \frac{\partial}{\partial \rho} 
\left( \frac{\rho'}{\rho} \right)^{|m|} \right] \right|_{\rho = \rho'}
- \left[ \left. \rho \, \frac{\partial}{\partial \rho} 
\left( \frac{\rho}{\rho'} \right)^{|m|} \right] \right|_{\rho = \rho'}
\right) 
= C \, \left(-|m| - |m|\right) \, \rho \, \frac{\rho^{|m|}}{\rho^{|m|+1}}
= \frac{1}{2 \pi} \,,
\end{equation}
with the result
\begin{equation}
\label{derEND}
C = -\frac{1}{2 \pi} \, \frac{1}{2 |m|} \,.
\end{equation}

The case $m=0$ requires special treatment. One sets
\begin{equation}
\label{ansatzD}
g_m(\rho, \rho') = D \, \ln\left( \frac{\rho_>}{L} \right) \,,
\end{equation}
because this term matches the asymptotic limit of 
Eq.~\eqref{gf} for $\rho \to \infty$, $\rho' \to 0$.
In this case, Eq.~\eqref{radcond} translates into the condition
\begin{equation}
D \, \left( 
\left. \left[ \rho \, \frac{\partial}{\partial \rho} 
\ln\left( \frac{\rho}{L} \right) 
\right] \right|_{\rho = \rho'}
- \left[ \left. \rho \, \frac{\partial}{\partial \rho} 
\ln\left( \frac{\rho'}{L} \right)
\right] \right|_{\rho = \rho'}
\right) 
= D - 0 = \frac{1}{2 \pi} \,,
\end{equation}
with the result $D = 1/(2 \pi)$.
Adding the terms for $m=0$ and $m \neq 0$, one has
\begin{align}
\label{beautiful}
g(\vec \rho - \vecrhoprime) 
=& \; \frac{1}{2 \pi} 
\ln\left( \frac{\rho_>}{L} \right)
- \sum_{\scriptsize \begin{array}{c} m\!=\!-\infty\\ m\neq 0\end{array}}^\infty 
\frac{1}{4 \pi |m|} 
\left( \frac{\rho_< }{\rho_>} \right)^{|m|} 
\ee^{\ii \, m \, (\varphi - \varphi')} 
\nonumber\\[0.1133ex]
=& \; \frac{1}{2 \pi} \, \left[ \ln\left( \frac{\rho_>}{L} \right)
- \sum_{m=1}^\infty \frac{1}{m} 
\left( \frac{\rho_< }{\rho_>} \right)^{m} 
\cos\left( m \, (\varphi - \varphi') \right) \right].
\end{align}

A numerical check of this relations is successful.
For $\vec\rho = 0.2 \, \hat{\rm e}_x + 0.1 \, \hat{\rm e}_y$
and $\vecrhoprime = 1.1 \, \hat{\rm e}_x + 1.5 \, \hat{\rm e}_y$,
{\color{blue} \fbox{and $L = 10.7$}},
the expression in Eq.~\eqref{gf} evaluates to
\begin{equation}
T_1 = g(\vec \rho - \vecrhoprime) =
\frac{1}{2 \pi} \, \ln\left( \frac{| \vec\rho - \vecrhoprime |}{L} \right) =
-0.296159 \,.
\end{equation}
while the $m=0$ term from Eq.~\eqref{beautiful} is 
\begin{equation}
T_2 = \frac{1}{2 \pi} \, 
\ln\left( \frac{\rho_>}{L} \right) = -0.278459 \,.
\end{equation}
Adding the sum over the nonzero $m$,
one obtains
\begin{equation}
\label{success}
T_3 = - \sum_{m=1}^\infty \frac{1}{2 \pi m} 
\left( \frac{\rho_< }{\rho_>} \right)^{m} 
\cos[ m \, (\varphi - \varphi') ] = 
-0.017700.
\end{equation}
We have checked the equality $T_1 = T_2 + T_3$ for a number of 
example cases. It is interesting to note that Eq.~\eqref{beautiful}
does not seem to have appeared in the literature before.

%
%
\section{Three--Dimensional Case}

Let $\vec r$ and $\vec r'$ denote coordinate vectors in 
three-dimensional space. It is well known that 
\begin{equation}
\label{g3D}
g(\vec r - \vec r') = 
-\frac{1}{4 \pi} \, \frac{1}{| \vec r - \vec r'| } \,,
\qquad
g(\vec k) = -\int \dd^3 r \,
\ee^{-\ii \vec k \cdot (\vec r - \vec r')} \,
\frac{1}{4 \pi} \, \frac{1}{|\vec r - \vec r'|} =
-\frac{1}{\vec k^2} \,,
\end{equation}
fulfills the Poisson equation
\begin{equation}
\vec\nabla^2 g(\vec r - \vec r') =
\delta^{(3)}( \vec r - \vec r' ) \,.
\end{equation}
The well-known expansion into (three-dimensional) 
spherical harmonics reads as follows,
\begin{equation}
\label{eq2}
g(\vec r - \vec r') = 
- \sum_{\ell =0}^{\infty }\sum_{m=-\ell }^{\ell }
\frac{1}{2\ell +1}
\frac{r_{<}^\ell}{r_>^{\ell + 1}} \;
Y_{\ell m}\left( \theta , \varphi \right) 
Y^\ast_{\ell m}\left( \theta', \varphi' \right) ,
\end{equation}
where $r_< = \min(r, r')$, $r_> = \max(r, r')$.
The Laplacian in three dimensions reads as 
\begin{equation}
\vec\nabla^2 = 
\frac{\partial^2}{\partial r^2}
+ \frac{2}{r} \frac{\partial}{\partial r} 
- \frac{\vec L^2}{r^2} \,,
\end{equation}
where $\vec L = -\ii \vec r \times \vec \nabla$.
The radial part of the Green function~\eqref{eq2} 
is assembled from homogeneous solutions of the 
radial equation, in much the same way 
as in the derivation extending from
Eq.~\eqref{derSTART} to Eq.~\eqref{derEND}.
The transformation from Cartesian to spherical coordinates is,
with $\vec r = \sum_{i=1}^3 x_i {\hat {\rm e}}_i$,
\begin{subequations}
\label{coords3}
\begin{align}
x_1 =& \; r \, \sin \theta \, \cos \varphi \,,
\\[0.1133ex]
x_2 =& \; r \, \sin \theta \, \sin \varphi \,,
\\[0.1133ex]
x_3 =& \; r \, \cos \theta \,,
\end{align}
\end{subequations}
The infinitesimal solid angle element is 
\begin{equation}
\dd^2 \Omega = \sin \theta \, \dd \theta \, \dd \varphi \,.
\end{equation}
The well-known spherical harmonics are given as
\begin{equation}
\label{defYlm}
Y_{\ell m}\left( \theta ,\varphi \right) =
\sqrt{\frac{2 \ell +1}{4\pi} \;
\frac{\left( \ell -m\right) !}{\left( \ell +m\right) !}} \;
P_{\ell}^{m}\left( \cos \left( \theta \right) \right) \;
\ee^{\ii m\varphi} \,,
\end{equation}
with the orthonormality and completeness properties
\begin{subequations}
\label{ortho}
\begin{align}
\int \dd^2 \Omega \;
Y_{\ell^{\prime} \, m^{\prime}}^{\ast}\left(\theta, \varphi \right) \,
Y_{\ell m}\left( \theta ,\varphi \right) =& \;
\delta_{\ell \ell ^{\prime }} \; \delta _{mm^{\prime }} \,,
\\[0.1133ex]
\sum_{\ell m} 
Y_{\ell\,m}\left( \theta,\varphi \right) 
Y^\ast_{\ell\,m}\left( \theta', \varphi'\right)
=& \; \frac{1}{\sin\theta}
\delta ( \theta - \theta' ) 
\delta ( \varphi -\varphi' ) .
\end{align}
\end{subequations}
The summation limits are $\ell = 0, \dots, \infty$ and 
$m = -\ell, \dots, \ell$.
The generating function for the Legendre polynomials~\cite{AbSt1972} is
\begin{equation}
\label{leggen}
P_\ell^{0}(x) = P_{\ell }(x) \,,
\qquad
\frac{1}{\sqrt{1 - 2 x t + t^2}} = 
\sum_{\ell = 0}^\infty P_\ell(x) \, t^\ell \,,
\end{equation}
while the associated Legendre polynomials are given by
\begin{equation}
\label{assocleg}
P_{\ell }^{m}\left( x\right) =
(-1)^m \, \left( 1-x^{2}\right)^{m\,/\,2} \;
\frac{\dd^{m} P_{\ell}\left( x\right) }{\dd x^{m}}\,.
\end{equation}
They have the property 
\begin{equation}
\label{legdiff}
\left( \frac{1}{\sin\theta} \frac{\partial}{\partial \theta} 
\sin\theta \frac{\partial}{\partial \theta} 
- \frac{m^2}{\sin^2 \theta} \right) P^m_\ell(\cos\theta) 
= -\ell \, (\ell + 1) \, P^m_\ell(\cos\theta) \,.
\end{equation}
These formulas are recalled with the notion of
clarifying the analogies with the four-dimensional case,
as will be done in the following.

%
%
\section{Four--Dimensional Case}

Let us denote four-dimensional vectors like
$\bm \xi$ and $\bm \xi'$ in bold face.
Just for clarity, we should stress that we are 
assuming a Euclidean metric.
The formula analogous to Eq.~\eqref{g3D} is
\begin{equation}
g( \bm\xi - \bm\xi' ) =
 -\frac{1}{4 \pi^2} \, \frac{1}{(\bm\xi - \bm\xi')^2} \,,
\qquad
g(\bm k) = -\int \dd^4 \xi \,
\ee^{-\ii \bm k \cdot (\bm\xi - \bm\xi')} \,
\frac{1}{4 \pi^2} \, \frac{1}{(\bm\xi - \bm\xi')^2} =
-\frac{1}{\bm k^2} \,,
\end{equation}
where $\bm \xi, \bm \xi' \in \mathds{R}^4$. 
The Green function $g(\bm \xi - \bm \xi')$ fulfills the 
equation
\begin{equation}
\bm\nabla^2 g(\bm\xi - \bm \xi') =
\delta^{(4)}( \bm \xi - \bm \xi' ) \,.
\end{equation}
The expansion into (four-dimensional) 
spherical harmonics~\cite{Sc1964} introduces an additional quantum 
number, which we denote as $n$,
and the analogue of Eq.~\eqref{eq2} is
\begin{equation}
\label{decomp}
g(\bm\xi - \bm\xi') = -\sum_{n \ell m}
\frac{1}{2 (n+1)} \, \frac{\xi_<^n}{\xi_>^{n+2}} \,
Y_{n \ell m}(\chi, \theta, \varphi) \,
Y^*_{n \ell m}(\chi', \theta', \varphi') \,,
\end{equation}
where the $Y_{n \ell m}(\chi, \theta, \varphi)$
are four-dimensional spherical harmonics,
and $\xi_< = \min( |\bm\xi|, |\bm\xi'| )$, 
and $\xi_> = \max( |\bm\xi|, |\bm\xi'| )$.
The summation limits are
$n=0, \dots,\infty$, 
$\ell = 0, \dots, n$, and 
$m = -\ell, \dots, \ell$.
The four-dimensional Laplacian is
\begin{equation}
\bm\nabla^2 =
\frac{\partial^2}{\partial \xi^2}
+ \frac{3}{\rho} \frac{\partial}{\partial \xi}
+ \frac{1}{\xi^2} 
\left( \frac{1}{\sin^2 \chi}       
\frac{\partial}{\partial \chi}
\sin^2 \chi \frac{\partial}{\partial \chi} -
\frac{\vec L^2}{\sin^2 \chi} \right),
\end{equation}
and the radial part of the decomposition~\eqref{decomp} 
is assembled from homogeneous solutions of the 
radial component of the four-dimensional Laplacian.
The transformation from Cartesian to spherical coordinates is,
with $\bm\xi = \sum_{i=1}^4 x_i \; {\hat {\bm e}}_i$,
\begin{subequations}
\label{coords4}
\begin{align}
x_1 =& \; r  \, \cos \varphi \, \sin \theta \, \sin \chi,
\\[0.1133ex]
x_2 =& \; r  \, \sin \varphi \, \sin \theta \, \sin \chi,
\\[0.1133ex]
x_3 =& \; r \, \cos \theta \, \sin \chi,
\\[0.1133ex]
x_4 =& \; r \, \cos \chi \,.
\end{align}
\end{subequations}
The infinitesimal solid angle element is 
\begin{equation}
\dd^3 \Omega = \sin^2 \chi \, \sin \theta \, \dd \theta \dd \varphi \,.
\end{equation}
The four-dimensional 
spherical harmonics can be given in terms of 
the analogue of Eq.~\eqref{defYlm} as 
\begin{equation}
\label{defYnlm}
Y_{n \ell m}(\chi, \theta, \varphi) 
= \sqrt{ \frac{2}{\pi} } \,
\sqrt{ \frac{(n+1) \, (n-\ell)!}{(n+\ell+1)!} } \;
Q^{\ell}_n(\cos\chi) \; Y_{\ell m}(\theta, \varphi) \,,
\end{equation}
with the orthonormality and completeness properties
\begin{subequations}
\label{ortho4}
\begin{align}
\int \dd^3 \Omega \,
Y_{n' \ell' m'}^{\ast}\left(\chi, \theta, \varphi \right) \,
Y_{n \ell m}\left( \chi, \theta, \varphi \right) =& \;
\delta_{n n'} \, 
\delta_{\ell \ell'} \; \delta_{m m'} \,,
\\[0.1133ex]
\sum_{n=0}^\infty \sum_{\ell=0}^n \sum_{m = -\ell}^\ell
Y_{n \, \ell\,m}\left( \chi, \theta, \varphi \right) \,
Y^\ast_{n \, \ell\,m}\left( \chi', \theta', \varphi'\right)
=& \; \frac{1}{\sin^2 \chi \, \sin\theta}
\delta \left( \chi - \chi' \right) \,
\delta \left( \theta - \theta' \right) \,
\delta \left( \varphi -\varphi' \right) \,.
\end{align}
\end{subequations}
The generating function for the Gegenbauer-type $Q$ polynomials is
an analogue of Eq.~\eqref{leggen},
\begin{equation}
\label{polyQ}
Q_n^{0}(x) = Q_n(x) \,,
\qquad
\frac{1}{1 - 2 x t + t^2} = 
\sum_{n = 0}^\infty Q_n(x) \, t^\ell \,.
\end{equation}
The associated Gegenbauer--type polynomials can be defined
in complete analogy with Eq.~\eqref{assocleg},
\begin{equation}
\label{assocQ}
Q_n^\ell\left( x\right) =
(-1)^\ell \, \left( 1 - x^2\right)^{\ell/2} \;
\frac{\dd^\ell Q_n\left( x\right) }{\dd x^\ell}\,.
\end{equation}
They have a property analogous to Eq.~\eqref{legdiff},
\begin{equation}
\left( \frac{1}{\sin^2 \chi}       
\frac{\partial}{\partial \chi}
\sin^2 \chi \frac{\partial}{\partial \chi} -
\frac{\ell \, (\ell + 1)}{\sin^2 \chi} \right) Q^\ell_n(\cos\chi) 
= - n \, (n + 2) \; Q^\ell_n(\cos\chi) \,.
\end{equation}
Steps toward a unified treatment of the four-dimensional 
spherical harmonics were made in Ref.~\cite{TCCA2007},
but it appears that the normalization prefactor
in Eq.~\eqref{defYnlm} was not given in explicit form.
The connection to the usual associated Gegenbauer polynomials
$C_{n}^{\ell}(x)$ 
(in the canonical form, see Ref.~\cite{AbSt1972}) is found as
\begin{subequations}
\begin{align}
Q_n(x) =& \; C_n^1(x) \,,
\\[0.1133ex]
Q_n^\ell(x) =& \; (-1)^n \, 2^\ell \, \ell! \, (1 - x^2)^{\ell/2} \,
C_{n-\ell}^{\ell + 1}(-x) \,.
\end{align}
\end{subequations}
Finally, we should mention the addition theorem
\begin{equation}
\sum_{n \ell m}
Y_{n \ell m}(\chi, \theta, \varphi) \,
Y^*_{n \ell m}(\chi', \theta', \varphi') =
\frac{n+1}{2 \pi^2} \,
Q_n( \bm x \cdot \bm x')  \,.
\end{equation}
Connections of these formulas to the hydrogen 
wave functions are discussed in the Appendix.

%
%
\section{Conclusions}

The most important formulas of this brief 
paper can be found in Eqs.~\eqref{beautiful},~\eqref{eq2}
and~\eqref{decomp}:
We derive [and in the case of Eq.~\eqref{eq2}, just recall]
the decomposition of the 
two-, three- and four-dimensional Green functions
of the Poisson equation into radial and 
angular parts. For $D=2$, only one ``quantum number''
is introduced, namely, the 
``magnetic'' (azimuthal) quantum number $m$;
for $D=3$, one has the orbital angular momentum
$\ell$ and its magnetic projection $m$,
while in $D=4$, a third additional quantum number has
to be introduced which can be associated with a
``principal'' quantum number $n$;
it is associated with the additional angular 
coordinate $\chi$ in four dimensions 
[see Eq.~\eqref{coords4}].
The latter interpretation is ramified by the fact 
that indeed, the momentum-space wave functions of the 
nonrelativistic hydrogen atom (for nuclear charge number $Z=1$) 
can be written as [cf.~p.~39 of Ref.~\cite{BeSa1957}]
%
%
%
\begin{equation}
\label{waveYnlm}
\psi_{n \ell m}(\vec p) =
(2 \pi)^{3/2} \,
\frac{4 [\hbar/(a_0 n)]^{5/2}}{([\hbar/(a_0 n)]^2 + \vec p^{\,2})^2} \,
Y_{(n -1) \, \ell m}(\chi, \theta, \varphi) \,,
\end{equation}
where
\begin{equation}
\cos\chi = 
\frac{[\hbar/(a_0 n)]^2 - \vec p^{\,2}}{[\hbar/(a_0 n)]^2 + \vec p^{\,2}} \,,
\end{equation}
and $\theta$ and $\varphi$ are the polar and azimuth 
angles of the unit vector in the momentum direction, 
i.e., in the direction of the unit vector $\hat p = \vec p / |\vec p|$.
The Bohr radius in $a_0 = \hbar/(\alpha m_e c)$, where 
$\alpha$ is the fine-structure constant, $m_e$ is the electron mass,
and $c$ is the speed of light.
These wave functions are normalized as 
$(2\pi)^{-3} \int \dd^3p \, | \psi_{n \ell m}(\vec p) |^2 = 1$.

In our treatment of the four-dimensional Green function,
we find it useful [see Eqs.~\eqref{polyQ} and~\eqref{assocQ}]
to define polynomials $Q_n(x)$,
and associated function $Q^\ell_n(x)$, which 
are related to, but not equal to, the Gegenbauer,
and associated Gegenbauer, polynomials~\cite{AbSt1972}.
Hence, we refer to them as ``Gegenbauer-type'' functions.
Analogies to the three-dimensional case (Legendre
and associated Legendre functions) are highlighted.
The most intriguing problem in the calculation 
of the two-dimensional Green function lies in the matching
of the $m = 0$ term from Eq.~\eqref{ansatzD} 
with the $m \neq 0$ term from Eq.~\eqref{derSTART};
the consideration of the asymptotic limit
$\rho_> \to \infty$ helps in finding the 
matching coefficients [see Eq.~\eqref{success}].

The angular-momentum decomposition~\eqref{beautiful}
for $D=2$ reveals that the dominant logarithmic 
term in the interaction of vortices in the 
two-dimensional sine--Gordon model is exclusively 
due to $S$-wave interactions.
The result might become useful as one tries 
to augment previous studies on high-$T_c$ 
Josephson-coupled, and magnetically coupled 
superconductors~\cite{NaEtAl2007jpcm1,NaEtAl2007jpcm2}
by the inclusion of higher-order derivative terms~\cite{JeNaNaPrep}.

%
%
\section*{Acknowledgments}

This research has been supported by the
National Science Foundation (Grants PHY--1403973 and
PHY--1710856) and by the Missouri Research Board.

\appendix

%
%
\section{Remarks on Schwinger's Derivation}

In this appendix, we provide the clarification of three points
which we found to be in need of some further explanation,
in regard to Schwinger's derivation~\cite{Sc1964,Li1968,Li1989} of the 
Schr\"{o}dinger--Coulomb Green function,
which is based on the $SO(4)$ symmetry.
First, one may observe that a certain
prefactor in the definition of the 
Schr\"{o}dinger--Coulomb Green function may be in need 
of a reconsideration.
Namely, if we assume that 
the defining equation of the Schr\"{o}dinger--Coulomb
Green function in coordinate space is
\begin{equation}
\left( \frac{\vec p^{\,2}}{2 m} + V - E \right) \, G(\vec r, \vec r')
= \delta^{(3)}(\vec r - \vec r') \,,
\end{equation}
where $V = -\frac{Z\alpha}{r}$ is the Coulomb potential,
then the defining equation of the momentum-space Green function incurs a 
prefactor $(2 \pi)^3$, in comparison to Ref.~\cite{Sc1964}.
The following conventions for the Fourier transforms 
\begin{equation}
f(\vec p) = \int \dd^3 r \, \ee^{-\ii \vec k \cdot \vec r} \, f(\vec r) \,,
\quad
f(\vec r) = \int \frac{\dd^3 p}{(2\pi)^3} \, \ee^{\ii \vec p \cdot \vec r} \, 
f(\vec p) \,,
\end{equation}
with an ``asymmetric'' distribution of the factors $2 \pi$,
are almost universally adopted in the physical literature.
With $\hbar = c = \epsilon_0 = 1$, 
the Coulomb potential, in momentum space, is
$V(\vec p -\vec p') = -\frac{4 \pi Z \alpha}{( \vec p - \vec p')^2}$.
The defining equation for the Green function
thus becomes, in momentum space,
\begin{equation}
\left( \frac{\vec p^{\,2}}{2 m} - E \right) \, G(\vec p, \vec p')
- \int \frac{\dd^3 p''}{(2 \pi)^3} \,
\frac{4 \pi Z \alpha}{( \vec p - \vec p'')^2} \, G(\vec p'', \vec p') 
= (2 \pi)^3 \, \delta^{(3)}(\vec p - \vec p') \,.
\end{equation}
The factor $(2\pi)^3$ is not present in the first (unnumbered)
equation of Ref.~\cite{Sc1964}.

The fifth (unnumbered) equation of Ref.~\cite{Sc1964} contains 
two nontrivial identities.
It is useful to derive the equation
$\dd^3 \Omega = \frac{\dd^3 \xi}{| \xi_0 |} $
for the area element on the three-dimensional unit sphere, 
embedded in four-dimensional space.
Here, one should remember that the three-dimensional 
components of the four-dimensional vector $(\xi_0, \vec \xi)$
may have varying magnitude, but one considers
them, according to Ref.~\cite{Sc1964},
on the four-dimensional unit sphere
$\xi_0^2 + \vec \xi^{\,2} = 1$.
One needs to remember that 
the appropriate generalization to the three-dimensional 
``surface'' of a manifold embedded into four-dimensional space is,
with $x = x(t_1, t_2, t_3)$, $y = y(t_1, t_2, t_3)$,
$z = z(t_1, t_2, t_3)$, $a = a(t_1, t_2, t_3)$ 
being the fourth coordinate,
where $a$ is the fourth coordinate,
\begin{equation}
\dd^3 \Omega = 
\left| {\rm det} 
\left( \begin{array}{cccc} 
\hat{\rm e}_x & \hat{\rm e}_y & \hat{\rm e}_z & \hat{\rm e}_a \\[1ex]
\frac{\partial x}{\partial t_1} &
\frac{\partial y}{\partial t_1} &
\frac{\partial z}{\partial t_1} & 
\frac{\partial a}{\partial t_1} \\[1ex]
\frac{\partial x}{\partial t_2} &
\frac{\partial y}{\partial t_2} &
\frac{\partial z}{\partial t_2} &
\frac{\partial a}{\partial t_2} \\[1ex]
\frac{\partial x}{\partial t_3} &
\frac{\partial y}{\partial t_3} &
\frac{\partial z}{\partial t_3} &
\frac{\partial a}{\partial t_3} 
\end{array} \right) \right| \, \dd t_1 \, \dd t_2 \, \dd t_3 \,.
\end{equation}
One calculates first the four-dimensional vector 
described by the determinant,
and then calculates its vector modulus.
The three-dimensional unit sphere, embedded 
in four-dimensional space, can be 
interpreted as a three-dimensional manifold,
parameterized by the coordinates 
$x = \xi_x = t_1$, $y = \xi_y = t_2$, and 
$z = \xi_z = t_3$,
while $a = \xi_a = \sqrt{1 - \xi_x^2 - \xi_y^2 - \xi_z^2}$.
One finds that 
\begin{equation}
\dd^3 \Omega = 
\frac{\dd \xi_x \, \dd \xi_y \, \dd \xi_z}%
{ \sqrt{1 - \xi_x^2 - \xi_y^2 - \xi_z^2 } } =
\frac{\dd^3 \xi}{| \xi_0 |} \,.
\end{equation}
This finally 
shows the first identity in the fifth equation 
of Ref.~\cite{Sc1964}.
In order to show the second identity, one has to calculate a 
further Jacobian, transforming $\dd^3 \xi$ into $\dd^3 p$
(in the conventions of Ref.~\cite{Sc1964},
keeping the zeroth (or fourth) component $X = p_0$ of the four-dimensional
(Euclidean) momentum constant.

There is a second nontrivial point which we found to be 
not very well explained in Ref.~\cite{Sc1964},
and it concerns the fourth unnumbered equation 
(from the bottom) on the second page of Ref.~\cite{Sc1964}.
The ``version of the expansion'' referred to in 
Ref.~\cite{Sc1964} necessitates the use of the 
following trick, which is to enter the angular-momentum 
expansion formula
given for $D(\bm \xi - \bm \xi') = -g(\bm \xi - \bm\xi')$ 
with the following values for $\bm\xi = \bm r_1$, and 
$\bm \xi' = \bm r_2$, as follows,
\begin{align}
\bm r_1 =& \; \rho \, \bm \xi \,,
\qquad
| \bm \xi | = 1 \,,
\qquad
| \bm r_1 | = \rho \,,
\qquad
\hat r_1 = \hat \xi \,,
\nonumber\\[0.1133ex]
\bm r_2 =& \; \bm\xi' \,,
\qquad
| \bm \xi' | = 1 \,,
\qquad
| \bm r_2 | = 1 \,,
\qquad
\hat r_2 = \hat \xi' \,,
\end{align}
with $0 < \rho < 1$,
so that $r_1 = \rho = r_<$, and $r_2 = 1 = r_>$.
The identity
\begin{align}
| \bm r_1 - \bm r_2 |^2
=& \; | \bm r_1 |^2 + | \bm r_2 |^2 - 2 \rho \, (\bm \xi \cdot \bm \xi')
\nonumber\\[0.1133ex]
=& \; ( 1 - \rho )^2 + \rho (\bm \xi - \bm \xi')^2 \,,
\end{align}
then follows, leading to
\begin{equation}
\frac{1}{4 \pi^2} \,
\frac{1}{( 1 - \rho )^2 + \rho (\bm\xi - \bm\xi')^2}
= \sum_{n \ell m}
\frac{\rho^n}{2 (n+1)} \,
Y_{n \ell m}(\chi, \theta, \varphi) \,
Y^*_{n \ell m}(\chi', \theta', \varphi') \,,
\end{equation}
which is the desired identity used in Ref.~\cite{Sc1964}.
We note that a representation of the 
$Y_{n \ell m}(\chi, \theta, \varphi)$ in terms
of elementary functions is not given in Ref.~\cite{Sc1964}.

The calculations of Ref.~\cite{Sc1964} 
culminate in the integral representation 
[see also Eq.~(B2) of Ref.~\cite{Pa1993}]
\begin{equation}
\label{G}
G(\vec p, \vec p') = 4 \pi m X^3 \,
\left( \frac{\ii \ee^{\ii \pi \nu}}{2 \sin(\pi \nu)} \right)
\int_1^{0^+} \dd \rho \, \rho^{-\nu} \,
\frac{\partial}{\partial \rho}
\frac{ ( 1 - \rho^2 )/\rho }{ \left[ X^{2} \, (\vec p - \vec p')^2  +
\frac{( 1 - \rho )^2}{4 \rho}
\, (X^2 + \vec p^{\,2}) \, (X^2 + \vec p'^{\,2})  \right]^2 } \,,
\end{equation}
where $E = - X^2/(2 m)$ is the energy argument of the 
Green function ($X = p_0$), 
and $\nu = Z \alpha m/\sqrt{ - 2 m E }$.
In comparison to Ref.~\cite{Sc1964},
the result for $G$ adds the prefactor $(2 \pi)^3$;
in the latter form,
it has been useful in Lamb shift calculations~\cite{Pa1993,JePa1996}.

The hydrogen wave functions in momentum space can be expressed as 
[cf.~Eq.~\eqref{waveYnlm}]
\begin{equation}
\label{psiSCHWINGER1}
\psi_{n \ell m}(\vec p) = \frac{16 \, \pi \, a_0^3 \, n^2}{Z^3} \,
\sqrt{ \frac{ (n-1-\ell)! }{ (n+\ell)! } } \;
\left(1 + \frac{n^2 \, a_0^2 \, \vec p^{\,2}}{Z^2} \right)^{-2} \,
Q^{\ell}_{n-1}\left(
\frac{1 - \frac{n^2 \, a_0^2 \, \vec p^{\,2}}{Z^2}}%
{1 + \frac{n^2 \, a_0^2 \, \vec p^{\,2}}{Z^2}}
\right) \;
Y_{\ell m}(\theta, \varphi) \,,
\end{equation}
where $a_0$ is the Bohr radius, and $Z$ is the nuclear charge number.
In comparison to p.~39 of \cite{BeSa1957},
we absorb the overall prefactor $(-1)^{n-1}$
into the global phase of the wave function.

\end{document}